\documentclass[my]{myour}
\usepackage{latexsym}
\usepackage{amssymb}
\usepackage{amsmath}
\usepackage{graphics}
\usepackage{epsf}
\usepackage[dvips]{graphicx}
\begin{document}
\title{Slope analysis for elastic nucleon-nucleon scattering}

\author{V.A. Okorokov\inst{}
\thanks{e-mail: VAOkorokov@mephi.ru;~Okorokov@bnl.gov}%
} 
\institute{Moscow Engineering Physics Institute (State
University), \\ Kashirskoe Shosse 31, 115409 Moscow, Russian
Federation}
\date{date April 23, 2009}
%
\abstract{The diffraction slope parameter is investigated for
elastic proton-proton and proton-antiproton scattering based on
the all available experimental data at low and intermediate
momentum transfer va\-lu\-es. Energy dependence of the elastic
diffraction slopes is approximated by various analytic functions.
The expanded "standard" logarithmic approximations with minimum
number of free parameters allow to describe experimental slopes in
all available energy range reasonably. The estimations of
asymptotic shrinkage parameter $\alpha'_{\cal{P}}$ were obtained
for various $|t|$ domains based on the all available experimental
data. Various approximations differ from each other both in the
low energy and very high energy domains. Predictions for
diffraction slope parameter are obtained for elastic proton-proton
scattering at NICA, RHIC and LHC energies, for proton-antiproton
elastic reaction in FAIR energy domain for various approximation
functions. \PACS{
      {13.75.Cs}{Nucleon-nucleon interactions}   \and
      {13.85.Dz}{Elastic scattering}
     } 
} 
\maketitle
\section{Introduction}
\label{intro} Elastic hadron-hadron scattering, the simplest type
of had\-ro\-nic collision process, remains one of the topical
theoretical problems in fundamental interaction physics at
present. Forward elastic scattering process is an excellent test
for some fundamental principles (unitarity, analyticity,
asymptotic theorems) of modern approaches. In the case of $pp$ and
$\bar{p}p$ elastic scattering, although many experiments have been
made over an extended range of initial energies and momentum
transfer, these reactions are still not well understood. One can
suggests that by the time the accelerator complexes like RHIC, LHC
etc. are operating, the interest to the soft physics increases
significantly. In the absence of a pure QCD description of the
elastic $pp / \bar{p}p$ and these large-distance scattering states
(soft diffraction), an empirical analysis based on
model-independent fits to the physical quantities involved plays a
crucial role \cite{Avila-EPJ-C47-171-2006}. Therefore, empirical
fits of energy dependencies of global scattering parameters have
been used as an important source of the model-independent
information. This approach for $\sigma_{tot}$ and $\rho$ was
recently used in \cite{Okorokov-arXiv-0711.2231}. The third
important quantity for nucleon elastic scattering is the slope
parameter. The nuclear slope $B$ for elastic scattering is of
interest in its own right. This quantity defined according to the
following equation:
\begin{equation}
B\left(s,t\right)=\frac{\textstyle
\partial}{\textstyle
\partial t}\left(\ln \frac{\textstyle \partial \sigma\left(s,t\right)}{\textstyle
\partial t}\right), \label{eq:Slope-def}
\end{equation}
is determined experimentally by fitting the differential cross
section $d\sigma / dt$ at some collision energy $\sqrt{s}$. On the
other hand the study of $B$ parameter is important, in particular,
for reconstruction procedure of full set of helicity amplitudes
for elastic nucleon scattering \cite{Okorokov-arXiv-0711.2231}. In
the last 20-30 years, high-energy $\bar{p}p$ colliders have
extended the maximum $\bar{p}p$ collision energy from $\sqrt{s}
\sim 20$ GeV to $\sqrt{s} \sim 2$ TeV, the RHIC facility allows to
obtain $pp$ data up to $\sqrt{s}=500$ GeV\footnote{According to
the RHIC Run plan the $pp$ data at highest energy $\sqrt{s}=500$
GeV have to be collected during the spring -- summer Run 2009.}.
As consequence, the available collection of $pp$ and $\bar{p}p$
slope data from literature has extended. The present status of
slope for elastic $pp$ and $\bar{p}p$ scattering is discussed over
the full energy domain. Predictions for further facilities are
obtained based on the available experimental data.
\section{Experimental slope energy dependence}
\label{sec:1} We have attempted to describe the energy behavior of
the elastic nuclear slopes for $pp$ and $\bar{p}p$ reactions. The
classical Pomeron theory gives in first approximation the
following expression for the differential cross section of elastic
scattering at asymptotically high energies:
\begin{equation*}
d\sigma / dt \propto s^{2\left(\alpha_{\cal{P}}(t)-1\right)},
\end{equation*}
where $\alpha_{\cal{P}}(t)$ is a Pomeron trajectory. If
$\alpha_{\cal{P}}(t)$ is linear function of momentum transfer,
i.e. $\alpha_{\cal{P}}(t)=\alpha_{\cal{P}}(0)+\alpha'_{\cal{P}}t$,
then for the slope parameter $B\left(s\right)$ at some $t$ using
the definition (\ref{eq:Slope-def}) one can obtain
\begin{equation*}
B\left(s\right) \propto 2\alpha'_{\cal{P}}\ln s.
\end{equation*}
Indeed the ensemble of experimental data on slopes for elastic
nucleon collisions can be approximated reasonably by many
phenomenological approaches, at least for $\sqrt{s} > 20$ GeV. But
models contradict the experimental data at lower energies and / or
phenomenological approaches have a significant number of free
parameters as usual. On the other hand it is apparent from
previous investigations that the experimental data on slope
parameters do not follow the straight lines at any initial
energies when plotted as function of $\ln s$. The new "expanded"
logarithmic parameterizations with small number of free parameters
have been suggested in
\cite{Okorokov-arXiv-0711.2231,VAOkorokov-arXiv-0811.0895-0811.3849}
for description of the elastic slope at all available energies.
Thus taking into account standard Regge parametrization and
quadratic function of logarithm from
\cite{Block-RevModPhys-57-563-1985} the following analytic
equations are under study here:
\begin{subequations}
\begin{eqnarray}
B\left(s,t\right)&=& B_{0}+2a_{1}\ln\left(s/s_{0}\right),
\label{eq:Fit-1} \\
B\left(s,t\right)&=&
B_{0}+2a_{1}\ln\left(s/s_{0}\right)+a_{2}\left
[\ln\left(s/s_{0}\right)\right]^{a_{3}},
\label{eq:Fit-2}\\
B\left(s,t\right)&=&
B_{0}+2a_{1}\ln\left(s/s_{0}\right)+a_{2}\left(s/s_{0}\right)^{a_{3}},
\label{eq:Fit-3}\\
B\left(s,t\right)&=&
B_{0}+2a_{1}\ln\left(s/s_{0}\right)+a_{2}\left
[\ln\left(s/s_{0}\right)\right]^{2}, \label{eq:Fit-4}
\end{eqnarray}
\end{subequations}
where $s_{0}=1$ GeV$^{2}$, in general case parameters $B_{0},~
a_{i}, i=1-3$ depend on range of $|t|$ which is used for
approximation. The function (\ref{eq:Fit-4}) is a special case of
(2b) at fixed value $a_{3}=2$. Additional terms in
(\ref{eq:Fit-2}) -- (\ref{eq:Fit-4}) take into account the
non-logarithmic part of the energy dependence of the elastic
nuclear slopes. Approximation function (\ref{eq:Fit-3}) is analogy
of parametrization of momentum slope dependence from
\cite{Burq-NPB-217-285-1983}. One can see the parametrization
(\ref{eq:Fit-3}) is compatible to first order with the Regge pole
model where the additional term represents the interference
between the Pomeron and secondary trajectories
\cite{Burq-NPB-217-285-1983}.

Most of experimental investigations as well as theoretical models
are focused on the diffraction region $|t| \simeq 0-0.5$
GeV$^{2}$. In this paper we study all available experimental data
for nuclear slope parameter up to $|t| \simeq 3.6$ GeV$^{2}$.
Experimental data are from
\cite{Lasinski-NPB-37-1-1972,Ambast-PRD-9-1179-1974,Jenni-NPB-129-232-1977,Fajardo-PRD-24-46-1981,Iwasaki-NPA-433-580-1985,DDG-url}.
The data sample consists of about 500 experimental points. The
ensemble of experimental data has been slightly specified and
improved in comparison with
\cite{VAOkorokov-arXiv-0811.0895-0811.3849}. In particular,
results obtained in \cite{Fajardo-PRD-24-46-1981} for form-factor
parametrization of $d\sigma / dt$ were included in the data sample
after detail additional investigation. The total number of
experimental points is equal $142 / 138$ for $pp / \bar{p}p$
scattering at low $|t|$, respectively. In the intermediate $|t|$
domain experimental data set is $136 / 69$ for $pp / \bar{p}p$
reaction respectively. Thus the experimental sample is
significantly larger than that for some early investigations
\cite{Okorokov-arXiv-0711.2231,Block-RevModPhys-57-563-1985,Burq-NPB-217-285-1983,Lasinski-NPB-37-1-1972,Block-PhysRep-436-71-2006}.
The careful analysis of data sample allows to suggest that the
influence of double counting in the experimental data is
negligible. It should be emphasized that the experimental data for
intermediate $|t|$ range were separated on two samples which
corresponded the various parametrization types for differential
cross-section, namely, linear, $\ln\left(d\sigma/dt\right) \propto
\left(-B|t|\right)$, and quadratic, $\ln\left(d\sigma/dt\right)
\propto \left(-B|t| \pm Ct^{2}\right)$, function. Here $B,C>0$ are
suggested. As known the measurements of nuclear slope, especially
at intermediate $|t|$ do not form a smooth set in energy, in
contrast to the situation for global scattering parameters $\rho$
and $\sigma_{tot}$, where there is a good agreement between
various group data \cite{Block-RevModPhys-57-563-1985}. Detailed
comparisons of slope data from different experiments are difficult
because the various experiments cover different $|t|$ ranges, use
various fitting procedures, treat systematic errors in different
ways, and, moreover, some experimental details are lost,
especially, for very early data. We have tried to use as much as
possible data for fitting from available samples. But some of the
$B$ values were not further used, either due to internal
inconsistencies in the fitting procedure, or as redundant in view
of a better determination at a nearby initial energy. Thus the
data samples for approximations are somewhat smaller because of
exclusion of points which, in particular, differ significantly
from the other experimental points at close energies. Critical for
a consistent determination of slope parameters is the choice of
the range $|t|_{\mbox{\small{min}}} \! \leq \! |t| \! \leq \!
|t|_{\mbox{\small{max}}}$ over which the fit of $d\sigma / dt$ is
performed. It seems both the mean value of
$|t|~\left(|\bar{t}|\right)$ and $|t|$-boundaries of corresponding
measurements are important for separation of experimental results
on different $|t|$ domains. Here the $\bar{|t|}$ are calculated
with taking into account approximations of experimental
$d\sigma/dt$ distributions instead of identifying of $|\bar{t}|$
with mean point of $|t|$-range as previously
\cite{VAOkorokov-arXiv-0811.0895-0811.3849}. Errors of
experimental points include available clear indicated systematic
errors added in quadrature to statistical ones. One need to
emphasize the systematic errors caused by the uncertainties of
normalization (total or/and differential cross sections) are not
taken into account if these uncertainties are not included in the
systematic errors in the original papers.

Let us describe the fitting algorithm in more detail. We use the
fitting procedure with standard likelihood function for this
investigation of nuclear slope parameter. In accordance with
\cite{Block-PhysRep-436-71-2006} let us define the quantity
\begin{equation}
\Delta \chi^{2}_{i}\left(s_{i};\vec{\alpha}\right) \equiv
\left(\frac{\textstyle
B_{m}^{i}-B\left(s_{i};\vec{\alpha}\right)}{\textstyle
\sigma_{i}}\right)^{2},\label{eq:Chi-def}
\end{equation}
where $B_{m}^{i}$ is the measured value of nuclear slope at
$s_{i}$, $B\left(s_{i};\vec{a}\right)$ is the expected value from
the one of the fitting functions under consideration, $\sigma_{i}$
is the experimental error of the $i$-th measurement. The
parameters $\alpha_{j}$ are given by the $N$-dimensional vector
$\vec{\alpha}=\left\{\alpha_{1},...,\alpha_{N}\right\}$. Our
fitting algorithm is some similar to the  "sieve" algorithm from
\cite{Block-PhysRep-436-71-2006} with following modification. We
reject the points which \textit{a priori} differ significantly
from other experimental data at close energies. The step allows us
to get a first estimation of $\chi^{2}/\mbox{n.d.f.}$ with minimum
number of rejected points. The fit quality is improved at the next
steps consequently. As indicated above smoothness of experimental
slope energy dependence differs significantly for data samples in
various $|t|$-domains and for various parameterizations of
$d\sigma / dt$ (see below). The $\Delta
\chi^{2}_{i}\left(s_{i};\vec{\alpha}\right)$ absolute value can be
large for one data sample but it can be acceptable for another
sample at the same time. Therefore we suggest to use the relative
quantity
\begin{equation}
n_{\chi}=\frac{\textstyle \Delta
\chi^{2}_{i}\left(s_{i};\vec{\alpha}\right)}{\textstyle
\chi^{2}/\mbox{n.d.f.}} \label{eq:Chi-rel}
\end{equation}
in order to reject the outliers (points far off from the fit curve
to the data sample) instead of constant cut value $\Delta
\chi^{2}_{i}\left(s_{i};\vec{\alpha}\right)_{\mbox{\small{max}}}$
from the "sieve" algorithm \cite{Block-PhysRep-436-71-2006}. One
needs to emphasize the fit function with best
$\chi^{2}/\mbox{n.d.f.}$ among (\ref{eq:Fit-1}) --
(\ref{eq:Fit-4}) is used for expected value calculation in
(\ref{eq:Chi-def}) at the each algorithm step. The points with
$n_{\chi} \geq n_{\chi}^{\mbox{\small{max}}}$ are excluded from
future study in our algorithm, where the
$n_{\chi}^{\mbox{\small{max}}}$ is some empirical cut value. The
conventional fit is made to the new "sifted" data sample. We
consider the estimates of fit parameters as the final results if
there are no excluded points for present data sample. We use the
one value $n_{\chi}^{\mbox{\small{max}}}=2$ for all data samples
considered in this paper below. The fraction of excluded points is
about $2\%$ for $pp$ as well as for $\bar{p}p$ elastic scattering
for low $|t|$ domain. The maximum relative amount of rejected
points is about $3\% / 12\%$ for linear $\ln d\sigma / dt$
parametrization and $6\% / 15\%$ for quadratic one at intermediate
$|t|$ values for $pp / \bar{p}p$ scattering respectively.
\subsection{Low $|t|$ domain}
\label{sec:2.1} 
\begin{figure}[h]
\resizebox{0.5\textwidth}{!}{%
  \includegraphics[width=8.0cm,height=8.0cm]{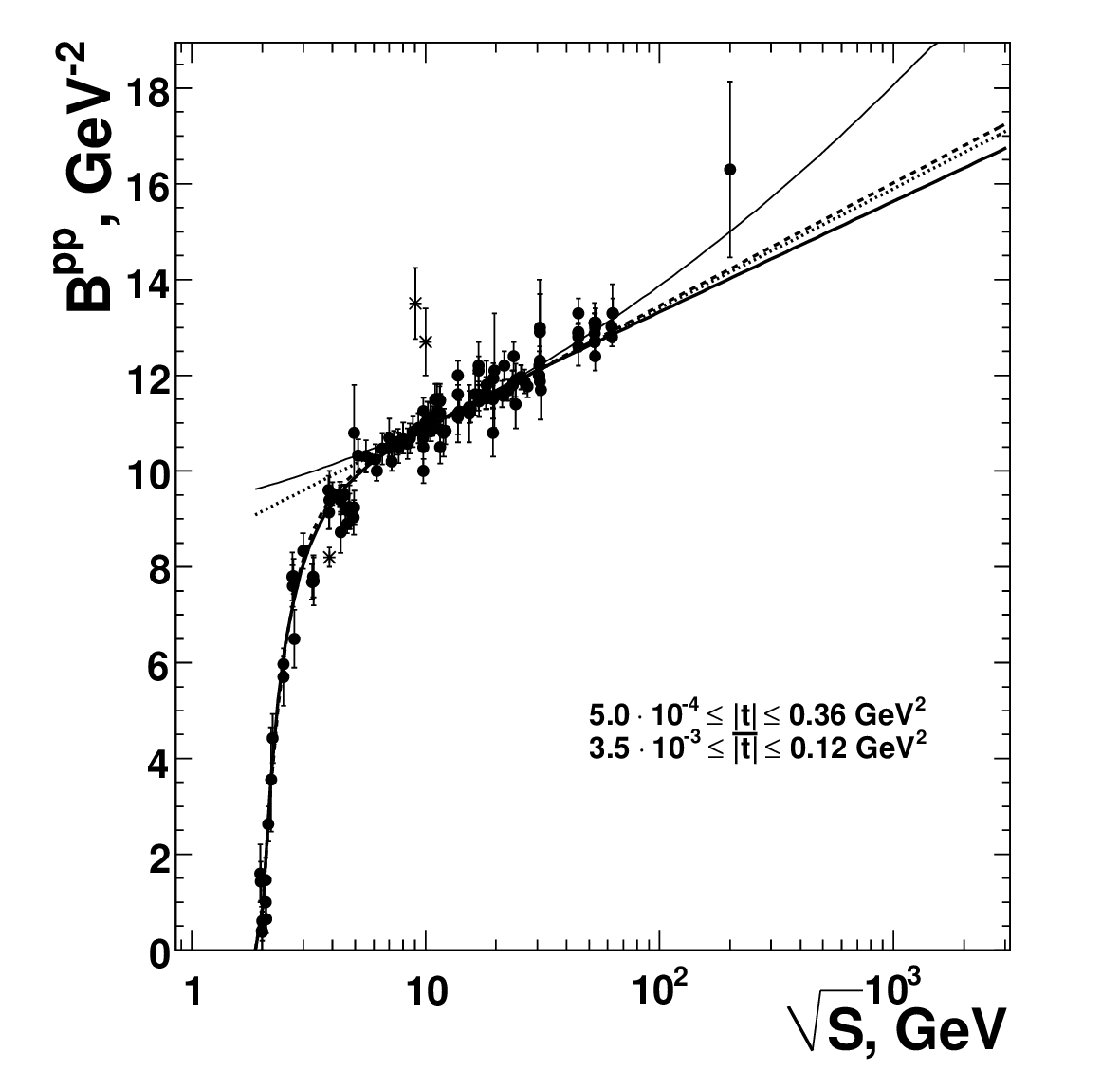}
} \caption{Energy dependence of the elastic slope parameters for
proton-proton scattering for low $|t|$ domain. Experimental fitted
points are indicated as {\large$\bullet$}, unfitted points are
indicated as {\large$*$}. The curves correspond to the fitting
functions as following: (\ref{eq:Fit-1}) -- dot, (\ref{eq:Fit-2})
-- thick solid, (\ref{eq:Fit-3}) -- dot-dashed, (\ref{eq:Fit-4})
-- thin} \label{fig:1}
\end{figure}

\begin{figure}
\resizebox{0.5\textwidth}{!}{%
  \includegraphics[width=8.0cm,height=8.0cm]{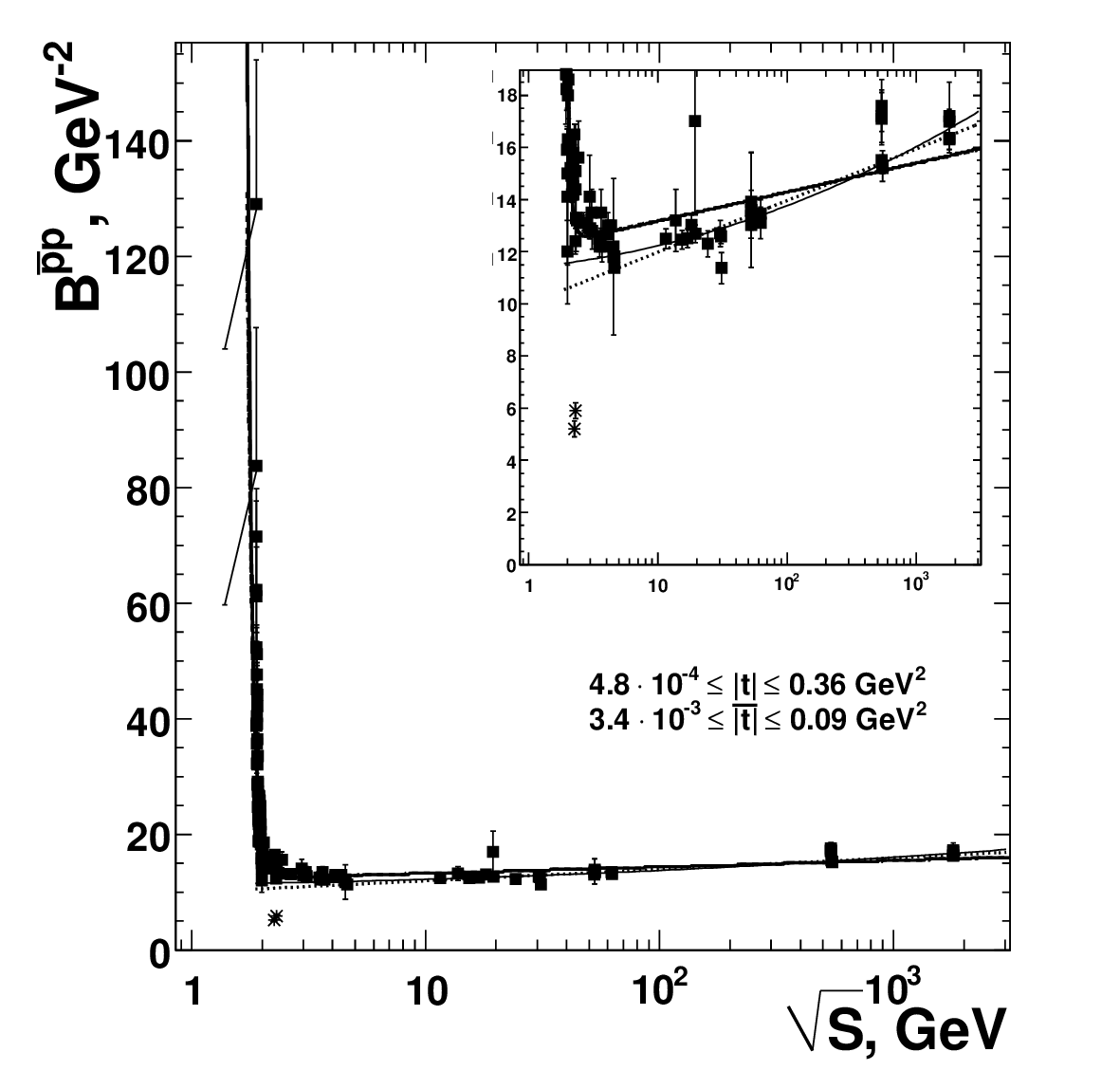}
} \caption{Energy dependence of the elastic slope parameters for
antiproton-proton scattering for low $|t|$ domain. Experimental
fitted points are indicated as $\blacksquare $, unfitted points
are indicated as {\large$*$}. The correspondence of curves to the
fit functions is the same as previously. The inner picture shows
the experimental data and fits at the same scale as well as for
fig.1}\label{fig:2}
\end{figure}

The energy dependence for experimental slopes and
corresponding fits by (\ref{eq:Fit-1}) -- (\ref{eq:Fit-4}) are
shown at Fig.\ref{fig:1} and Fig.\ref{fig:2} for $pp$ and
$\bar{p}p$ correspondingly. The fitting parameter values are
indicated in Table 1.
One can see that the fitting functions (\ref{eq:Fit-1}),
(\ref{eq:Fit-4}) describe the $pp$ (Fig.\ref{fig:1}) and
$\bar{p}p$ (Fig.\ref{fig:2}) experimental data statistically
acceptable only for $\sqrt{s} \geq 5$ GeV. Additional study
demonstrate that the extension of approximation domain to the
lower energies for parameterizations (\ref{eq:Fit-1}) and
(\ref{eq:Fit-4}) results in significant increasing of
$\chi^{2}/\mbox{n.d.f.}$ for $\bar{p}p$ data. Thus these fit
functions allow to get a statistically acceptable fit qualities
only at $\sqrt{s} \geq 5$ GeV for $\bar{p}p$ scattering as well as
for elastic $pp$ reaction. The parameter $a_{1}$ for
Regge-inspired function (\ref{eq:Fit-1}) is close to estimation
for Pomeron parameter $\alpha'_{\cal{P}} \approx 0.25$ GeV$^{-2}$
for $\bar{p}p$ data but this parameter is some larger than
$\alpha'_{\cal{P}}$ estimate for proton-proton data. The
$a_{1}^{pp}$ value for (\ref{eq:Fit-1}) from Table 1 is equal
within the errors the early experimental estimations of "true"
Pomeron shrinkage parameter \cite{Giacomelli-PR-23-123-1976}. The
parameter $a_{1}$ for function (\ref{eq:Fit-2}) is equal within
errors to the above estimation for Pomeron theory parameter
$\alpha'_{\cal{P}}$ for $pp$ data. The value of $a_{1}^{pp}$
parameter for fitting function (\ref{eq:Fit-3}) agrees with early
value for similar fit of slope momentum dependence
\cite{Burq-NPB-217-285-1983} but the present result is more
precise than previous one. The RHIC point for proton-proton
collisions has a large error and can't discriminate the
approximations. Fitting functions (\ref{eq:Fit-2}),
(\ref{eq:Fit-3}) allow us to describe experimental data at all
energies with reasonable fit quality for $pp$. The functions
(\ref{eq:Fit-1}) -- (\ref{eq:Fit-3}) are close to each other at
$\sqrt{s} \geq 5$ GeV, especially, the Regge model approximation
and (\ref{eq:Fit-3}) fit. It seems the ultra-high energy domain is
suitable for separation of various parameterizations. The
qualities of (\ref{eq:Fit-2}), (\ref{eq:Fit-3}) approximations for
$\bar{p}p$ elastic scattering data are much poorer because of very
sharp behavior of experimental data near the low energy boundary.
But one can see that the functions (\ref{eq:Fit-2}),
(\ref{eq:Fit-3}) agree with experimental points at qualitative
level and (very) close to each other for all energy range. The fit
quality is some better for function (\ref{eq:Fit-2}) than that for
parametrization (\ref{eq:Fit-3}) both for $pp$ and $\bar{p}p$
data. Additional study of antiproton-proton data shows that the
increasing of low boundary of fit range
$\left(s_{\mbox{\small{min}}}\right)$ leads to the better fit
quality for functions (\ref{eq:Fit-2}), (\ref{eq:Fit-3}) but at
the same time, obviously, to the lost of some low-energy
$\bar{p}p$ data. The fit quality changes dramatically at small
increasing of $s_{\mbox{\small{min}}}$ from low boundary value
$4m_{p}^{2}$ to 3.72 GeV$^{2}$. It was obtained
$\chi^{2}/\mbox{n.d.f.} \! \simeq \! 5.4 / 5.5$ for function
(\ref{eq:Fit-2}) / (\ref{eq:Fit-3}) respectively for fit range
$s_{\mbox{\small{min}}} \! \geq \! 3.72$ GeV$^{2}$. On the other
hand the data sample is about $75\%$ from maximum one in this
case. Thus it seems the $s_{\mbox{\small{min}}}=3.72$ GeV$^{2}$
one of the optimum values from point of view both fit quality and
closing to the threshold $s_{\mbox{\small{min}}}=4m_{p}^{2}$. The
average value of (\ref{eq:Chi-def}) for excluded points is equal
15.6 for $pp$ and 649.8 for $\bar{p}p$ data sample for
parametrization (2b).

\begin{table*}
\caption{Fitting parameters for slope energy dependence in low
$|t|$ domain} \label{tab:1}
\begin{center}
\begin{tabular}{lccccc}
\hline \multicolumn{1}{l}{Function} &
\multicolumn{5}{c}{Parameter} \\
\cline{2-6} \rule{0pt}{10pt}
 & $B_{0}$, GeV$^{-2}$ & $a_{1}$, GeV$^{-2}$ & $a_{2}$, GeV$^{-2}$ & $a_{3}$ & $\chi^{2}/\mbox{n.d.f.}$ \\
\hline
\multicolumn{6}{c}{proton-proton scattering} \\
\hline
(\ref{eq:Fit-1}) & $8.41 \pm 0.08$  & $0.271 \pm 0.007$ & --              & --               & $179/95$ \\
(\ref{eq:Fit-2}) & $8.73 \pm 0.12$  & $0.250 \pm 0.009$ & $-28 \pm 2$     & $-3.50 \pm 0.14$ & $349/135$ \\
(\ref{eq:Fit-3}) & $8.31 \pm 0.08$  & $0.279 \pm 0.007$ & $-188 \pm 23$   & $-2.24 \pm 0.09$ & $358/135$ \\
(\ref{eq:Fit-4}) & $9.3 \pm 0.3$    & $0.11 \pm 0.06$   & $0.03 \pm 0.01$ & --               & $171/94$ \\
\hline
\multicolumn{6}{c}{proton-antiproton scattering} \\
\hline
(\ref{eq:Fit-1}) & $10.0 \pm 0.2$   & $0.215 \pm 0.012$ & --                & --               & $32/27$ \\
(\ref{eq:Fit-2}) & $12.02 \pm 0.06$ & $0.121 \pm 0.006$ & $495 \pm 98$      & $-13.0 \pm 0.5$  & $1220/132$ \\
(\ref{eq:Fit-3}) & $12.06 \pm 0.05$ & $0.119 \pm 0.005$ & $\left(3.1 \pm 0.4\right) \cdot 10^{6}$& $-9.36 \pm 0.10$ & $1303/132$ \\
(\ref{eq:Fit-4}) & $11.4 \pm 1.0$   & $0.05 \pm 0.11$   & $0.017 \pm 0.011$ & -- & $29/26$ \\
\hline
\end{tabular}
\end{center}
\end{table*}
\begin{table*}
\caption{Predictions for nuclear slope based on the
parameterizations (\ref{eq:Fit-1}) -- (\ref{eq:Fit-4}) for low
$|t|$ domain} \label{tab:2}
\begin{center}
\begin{tabular}{lccccccccccc}
\hline \multicolumn{1}{l}{Fitting} &
\multicolumn{11}{c}{Facility energies, $\sqrt{s}$} \\
\cline{2-12} \rule{0pt}{10pt} function & \multicolumn{4}{c}{FAIR,
GeV} & \multicolumn{2}{c}{NICA, GeV} &
\multicolumn{2}{c}{RHIC, TeV} & \multicolumn{3}{c}{LHC, TeV} \\
\cline{2-12} \rule{0pt}{10pt}
 & 3 & 5 & 6.5 & 14.7 & 20 & 25 & 0.2 & 0.5 & 14 & 28 & 42$^*$\\
\hline (\ref{eq:Fit-1}) & -- & 11.38 & 11.61 & 12.31 & 11.66 &
11.90 & 14.15 & 15.15 & 18.76 & 19.51 & 19.95 \\
(\ref{eq:Fit-2}) & 12.57 & 12.80 & 12.93 & 13.32 & 11.67 & 11.91 &
14.02 & 14.94 & 18.28 & 18.97 & 19.37 \\
(\ref{eq:Fit-3}) & 12.59 & 12.83 & 12.95 & 13.34 & 11.65 & 11.90 &
14.22 & 15.25 & 18.96 & 19.74 & 20.19 \\
(\ref{eq:Fit-4}) & -- & 11.90 & 12.01 & 12.43 & 11.70 & 11.96 & 15.00 & 16.67 & 24.44 & 26.39 & 27.58 \\
\hline\multicolumn{10}{l}{$^*$\rule{0pt}{11pt}\footnotesize The
ultimate energy upgrade of LHC project \cite{Skrinsky-ICHEP2006}.}
\end{tabular}
\end{center}
\end{table*}
One can get predictions for nuclear slope parameter values for
some facilities based on the results shown above. The $B$ values
at low $|t|$ for different energies of FAIR, NICA, RHIC, and LHC
are shown in the Table 2. As expected the functions
(\ref{eq:Fit-2}) and (\ref{eq:Fit-3}) predicted very close slope
parameter values for FAIR. The approximation function
(\ref{eq:Fit-1}) and (\ref{eq:Fit-4}) can predict for $\sqrt{s}
\geq 5$ GeV only. Functions (\ref{eq:Fit-1}) -- (\ref{eq:Fit-3})
predict much smaller values for $B$ in high-energy $pp$ collisions
than (\ref{eq:Fit-4}) approximation. Perhaps, the future more
precise RHIC results will agree better with predictions based on
experimental data fits under study. Our prediction with
(\ref{eq:Fit-4}) function for ultimate energy of LHC agrees well
with early prediction for close SSC energy based only on slope
data in the region $5 < \sqrt{s} < 62$ GeV
\cite{Block-RevModPhys-57-563-1985}. It should be emphasized that
the fits (\ref{eq:Fit-1}) -- (\ref{eq:Fit-3}) of available
experimental data predict the slower increasing of $B$ with energy
than most of phenomenological models \cite{Kundrat-EDS-273-2007}.
The $B$ values predicted for LHC at $\sqrt{s}=14$ TeV by
(\ref{eq:Fit-1}) and (\ref{eq:Fit-3}) are most close to the some
model expectations
\cite{Block-PRD-60-054024-1999,Petrov-EPJ-C28-525-2003}. Moreover
one needs to emphasize that the model estimates at $\sqrt{s}=14$
TeV were obtained for $B\left(t=0\right)$ and the $t$-dependence
of slope shows the slight decreasing of $B$ at growth of momentum
transfer up to $|t| \approx 0.1-0.2$ GeV$^{2}$ in particular for
the models
\cite{Block-PRD-60-054024-1999,Petrov-EPJ-C28-525-2003}. Thus one
can expect the some better agreement between model estimations and
predicted values of $B$ from Table 2 for finite (non-zero) low
$|t|$ values. Possibly the saturation regime, Black Disk Limit,
will be reached at the LHC. The one of the models in which such
effects appear, namely, DDM predicts the slope $B\left(t=0\right)
\approx 23$ GeV$^{-2}$ at $\sqrt{s}=14$ TeV
\cite{Selyugin-EDS2007-279}.
\subsection{Intermediate $|t|$ domain}
\label{sec:2.1} As indicated above the situation is more
complicated for intermediate $|t|$ domain. Differential cross
section is approximated by linear, $\ln\left(d\sigma/dt\right)
\propto \left(-B|t|\right)$, or / and quad\-ra\-tic,
$\ln\left(d\sigma/dt\right) \propto \left(-B|t| \pm
Ct^{2}\right)$, function in various experiments, $|t|$ ranges used
for $d\sigma / dt$ approximations differ significantly etc. For
quadratic exponential parametrization the $B$ and $C$ parameters
are highly correlated by fits.

\begin{figure*}
\vspace*{1cm}
\includegraphics[width=17.0cm,height=17.0cm]{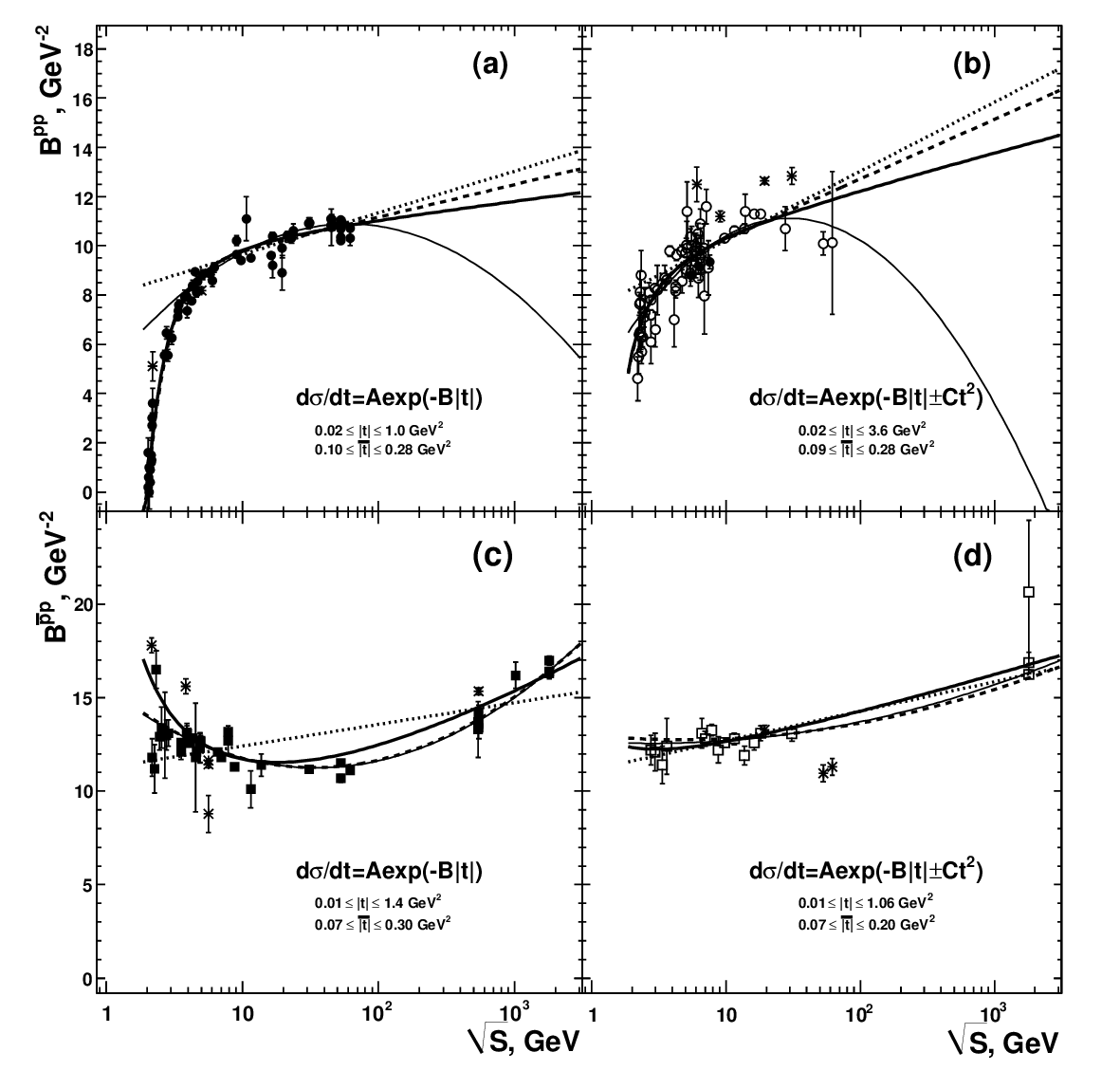}
\caption{Energy dependence of $B$ in proton-proton (a,b) and
proton-antiproton (c,d) elastic scattering for linear (a,c) and
quadratic (b,d) exponential parametrization of differential cross
section. Experimental points are indicated as close / open circles
(squares) for $pp~\left(\bar{p}p\right)$ for (a,c) / (b,d),
unfitted points are indicated as {\Large$*$}. The correspondence
of curves to the fit functions is the same as previously}
\label{fig:3}
\end{figure*}

Figure \ref{fig:3} shows the experimental data and corresponding
fits for slope parameter energy dependence at intermediate $|t|$
for $pp$ and $\bar{p}p$ elastic scattering. The Fig.\ref{fig:3}a
and Fig.\ref{fig:3}c correspond to the linear exponential
approximation of differential cross-section for $pp$ and
$\bar{p}p$ respectively. Experimental data obtained at quadratic
exponential fit of $d\sigma/dt$ and fitting functions
(\ref{eq:Fit-1}) -- (\ref{eq:Fit-4}) are presented at
Fig.\ref{fig:3}b for $pp$ and at Fig.\ref{fig:3}d for $\bar{p}p$
collisions. The fitting parameter values are indicated in Table
\ref{tab:3} for various interaction types and for different
$d\sigma/dt$ parameterizations.
Usually the fit qualities are poorer for intermediate $|t|$ values
than that for low $|t|$ range in $pp$ elastic collisions for
linear exponential parametrization of $d\sigma / dt$. The fitting
functions (\ref{eq:Fit-1}) and (\ref{eq:Fit-4}) agree with
experimental points qualitatively both for linear
(Fig.\ref{fig:3}a) and quadratic (Fig.\ref{fig:3}b) exponential
parameterizations of $d\sigma/dt$ for $\sqrt{s} \geq 5$ GeV only.
The "expanded" functions (\ref{eq:Fit-2}), (\ref{eq:Fit-3})
approximate experimental data at all energies reasonably with very
close fit qualities. But the (\ref{eq:Fit-2}) function shows a
very slow growth of slope parameter with energy increasing at
$\sqrt{s} \geq 100$ GeV (Fig.\ref{fig:3}a). It should be stressed
that the fitting function (\ref{eq:Fit-4}) predicts decreasing of
the nuclear slope in high energy domain. Such behavior is opposite
the other fitting function (\ref{eq:Fit-1}) -- (\ref{eq:Fit-3}).
In the case of linear exponential approximation of $d\sigma / dt$
mean value of characteristic (\ref{eq:Chi-rel}) is about 5.3 for
excluded $pp$ data points with (\ref{eq:Fit-2}) function and
$\overline{n_{\chi}} \simeq 2.6$ for points excluded from
$\bar{p}p$ fitted data sample for (\ref{eq:Fit-4}) fitting
function.

One can see that the $\bar{p}p$ experimental data admit the
approximation by (\ref{eq:Fit-4}) for all energy range but not
only for $\sqrt{s} \geq 5$ GeV. Indeed the fit quality for the
first case much better than for second one. Additional analysis
demonstrated just the same behavior of fit quality for function
(\ref{eq:Fit-1}) too. Thus $\bar{p}p$ experimental points from
linear exponential parametrization of differential cross-section
are fitted by (\ref{eq:Fit-1}) and (\ref{eq:Fit-4}) at all
energies not only for $\sqrt{s} \geq 5$ GeV. The parameter values
are shown in Table \ref{tab:3} for approximation by
(\ref{eq:Fit-1}), (\ref{eq:Fit-4}) of all available experimental
data. The $\bar{p}p$ data disagreement with Regge-inspired fitting
function very significantly (Fig.\ref{fig:3}c). Functions
(\ref{eq:Fit-3}) and (\ref{eq:Fit-4}) show a very close behavior
at all energies for $\bar{p}p$ data from linear parametrization of
$\ln d\sigma/dt$. These fitting functions have a better fit
quality than (\ref{eq:Fit-2}) but fits (\ref{eq:Fit-3}),
(\ref{eq:Fit-4}) are still statistically unacceptable. As
previously experimental data at Fig.\ref{fig:3}d allow the
approximation by (\ref{eq:Fit-1}) and (\ref{eq:Fit-4}) for all
energy range but not only for $\sqrt{s} \geq 5$ GeV. The fit
qualities are better in the first case of energy range and fitting
parameters are indicated in the Table \ref{tab:3} for this energy
range namely. As above the functions (\ref{eq:Fit-3}) and
(\ref{eq:Fit-4}) show a close fit quality which is some better
than this parameter for (\ref{eq:Fit-2}) fitting function. One can
see the fit qualities for (\ref{eq:Fit-1}) -- (\ref{eq:Fit-4}) are
better significantly for data from quadratic exponential
parametrization of differential cross-sections than for data from
linear exponential approximation of $d\sigma/dt$. Thus fitting
functions (\ref{eq:Fit-3}), (\ref{eq:Fit-4}) agree with data
points at quantitative level for quadratic (Fig.\ref{fig:3}d)
parametrization of proton-antiproton $\ln d\sigma/dt$ and these
fits are statistically acceptable. Excluded points are
characterized by $\overline{n_{\chi}} \simeq 17.9$ for $pp$ data
with (\ref{eq:Fit-2}) fitting function and by $\overline{n_{\chi}}
\simeq 12.1$ for $\bar{p}p$ data sample at (\ref{eq:Fit-4})
function.
\begin{table*}
\caption{Fitting parameters for slope energy dependence at
intermediate $|t|$} \label{tab:3}
\begin{center}
\begin{tabular}{lccccc}
\hline \multicolumn{1}{l}{Function} &
\multicolumn{5}{c}{Parameter} \\
\cline{2-6} \rule{0pt}{10pt}
 & $B_{0}$, GeV$^{-2}$ & $a_{1}$, GeV$^{-2}$ & $a_{2}$, GeV$^{-2}$ & $a_{3}$ & $\chi^{2}/\mbox{n.d.f.}$ \\
\hline
\multicolumn{6}{c}{proton-proton scattering, experimental data for $d\sigma/dt=A\exp\left(-B|t|\right)$} \\
\hline
(\ref{eq:Fit-1}) & $7.95 \pm 0.12$  & $0.184 \pm 0.009$ & --              & --               & $110/34$ \\
(\ref{eq:Fit-2}) & $9.7 \pm 0.3$    & $0.08 \pm 0.02$   & $-22.1 \pm 1.4$ & $-2.34 \pm 0.12$ & $240/60$ \\
(\ref{eq:Fit-3}) & $8.51 \pm 0.14$  & $0.144 \pm 0.010$ & $-71 \pm 5$     & $-1.49 \pm 0.06$ & $240/60$ \\
(\ref{eq:Fit-4}) & $4.9 \pm 0.6  $  & $0.73 \pm 0.10$   & $-0.09 \pm 0.02$& --             & $81/33$ \\
\hline
\multicolumn{6}{c}{proton-proton scattering, experimental data for $d\sigma/dt=A\exp\left(-B|t| \pm Ct^{2}\right)$} \\
\hline
(\ref{eq:Fit-1}) & $7.4 \pm 0.2$    & $0.31 \pm 0.03$   & --              & --               & $115/33$ \\
(\ref{eq:Fit-2}) & $9.6 \pm 2.4$    & $0.16 \pm 0.13$   & $-7.2 \pm 5.4$  & $-1.5 \pm 1.0$   & $227/62$ \\
(\ref{eq:Fit-3}) & $7.9 \pm 0.5$    & $0.26 \pm 0.05$   & $-23 \pm 16$    & $-1.5 \pm 0.5$   & $228/62$ \\
(\ref{eq:Fit-4}) & $4.1 \pm 0.9$    & $1.0 \pm 0.2$     & $-0.15 \pm 0.04$& --               & $102/32$ \\
\hline
\multicolumn{6}{c}{proton-antiproton scattering, experimental data for $d\sigma/dt=A\exp\left(-B|t|\right)$} \\
\hline
(\ref{eq:Fit-1}) & $11.25 \pm 0.06$ & $0.126 \pm 0.004$ & --              & --               & $1111/41$ \\
(\ref{eq:Fit-2}) & $\left(-5.0 \pm 0.9\right) \cdot 10^{3}$ & $0.64 \pm 0.02$ & $\left(5.1 \pm 0.9\right) \cdot 10^{3}$ & $\left(-1.5 \pm 0.3\right) \cdot 10^{-3}$ & $355/39$ \\
(\ref{eq:Fit-3}) & $-893 \pm 69$    & $5.9 \pm 0.3$   & $908 \pm 69$   & $\left(-1.44 \pm 0.06\right) \cdot 10^{-2}$ & $199/39$ \\
(\ref{eq:Fit-4}) & $15.46 \pm 0.15$ & $-0.59 \pm 0.02$  &$0.083 \pm 0.003$& --               & $197/40$ \\
\hline
\multicolumn{6}{c}{proton-antiproton scattering, experimental data for $d\sigma/dt=A\exp\left(-B|t| \pm Ct^{2}\right)$} \\
\hline
(\ref{eq:Fit-1}) & $11.1 \pm 0.2$   & $0.171 \pm 0.011$ & --                & --               & $17.3/15$ \\
(\ref{eq:Fit-2}) & $\left(-1\pm 3\right) \cdot 10^{3}$  & $0.26 \pm 0.06$   & $\left(1 \pm 3\right) \cdot 10^{3}$ & $\left(-1 \pm 4\right) \cdot 10^{-3}$ & $14.5/13$ \\
(\ref{eq:Fit-3}) & $\left(-2.2 \pm 1.8\right) \cdot 10^{2}$ & $1.7 \pm 0.8$ & $\left(2.3 \pm 1.8\right) \cdot 10^{2}$ & $-0.015 \pm 0.006$   & $13.2/13$ \\
(\ref{eq:Fit-4}) & $12.7 \pm 0.8$   & $-0.05 \pm 0.11$  &$0.023 \pm 0.011$& -- & $13.2/14$ \\
\hline
\end{tabular}
\end{center}
\vspace*{-0.4cm}
\end{table*}

From the quadratic exponential parametrization of differential
cross-section one may compute the local slope at a certain
$|t|$-value via the following relation
\begin{equation}
b\left(s,t\right)=B \pm C\ln|t|,~~~ B,C>0 \label{eq:b-Def}
\end{equation}
This characteristic can be useful for elastic scattering for study
of $d\sigma / dt$ in wider $|t|$ range. It is suggested $b \geq 0$
according to the definition (\ref{eq:b-Def}). The $b$-parameter is
named slope too, it is evaluated at $|t|=0.2$ GeV$^{2}$ usually.
One of the advantages of this characteristic is the expectation of
more smooth energy (momentum) dependence than that for
$B$-parameter discussed above. Indeed we have included the 100\%
of available experimental points in fitted sample for $pp$ elastic
scattering. But the number of points is some smaller than that for
$B$-parameter because of absent $C$-parameter values for some low
energy measurements from \cite{Lasinski-NPB-37-1-1972}. We
excluded one point at $\sqrt{s}=1.8$ TeV
\cite{Amos-PLB-247-127-1990} from fitted sample for $\bar{p}p$
elastic reaction because there are unacceptably large errors
(relative error is $\delta b \simeq 2.72$) for this point.

\begin{figure}
\resizebox{0.5\textwidth}{!}{%
  \includegraphics[width=8.0cm,height=8.0cm]{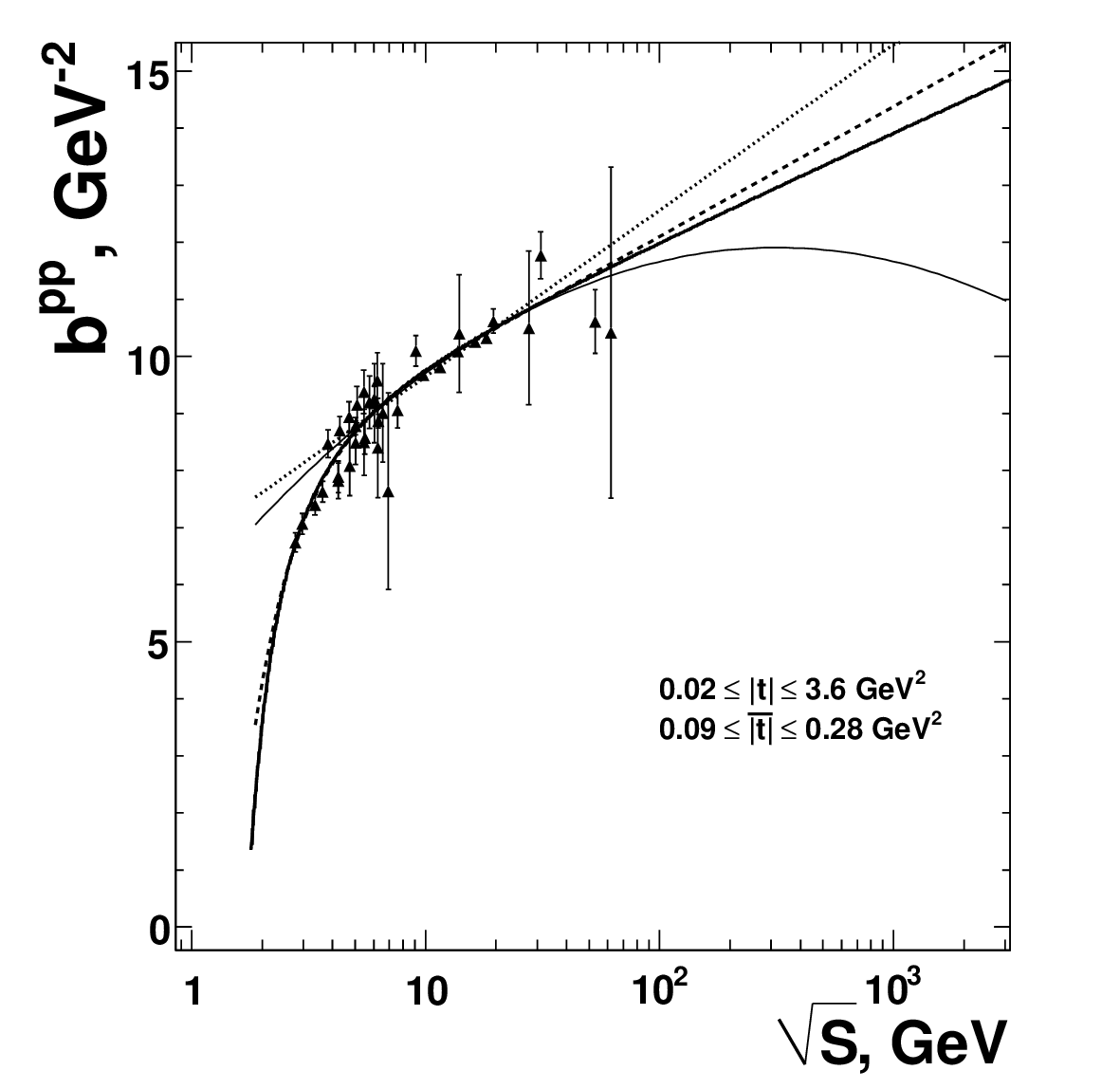}
} \caption{Energy dependence of the slope parameter $b$ at
$|t|=0.2$ GeV$^{2}$ for proton-proton scattering. The
correspondence of curves to the fit functions is the same as
previously}\label{fig:4}
\end{figure}

\begin{figure}
\resizebox{0.5\textwidth}{!}{%
  \includegraphics[width=8.0cm,height=8.0cm]{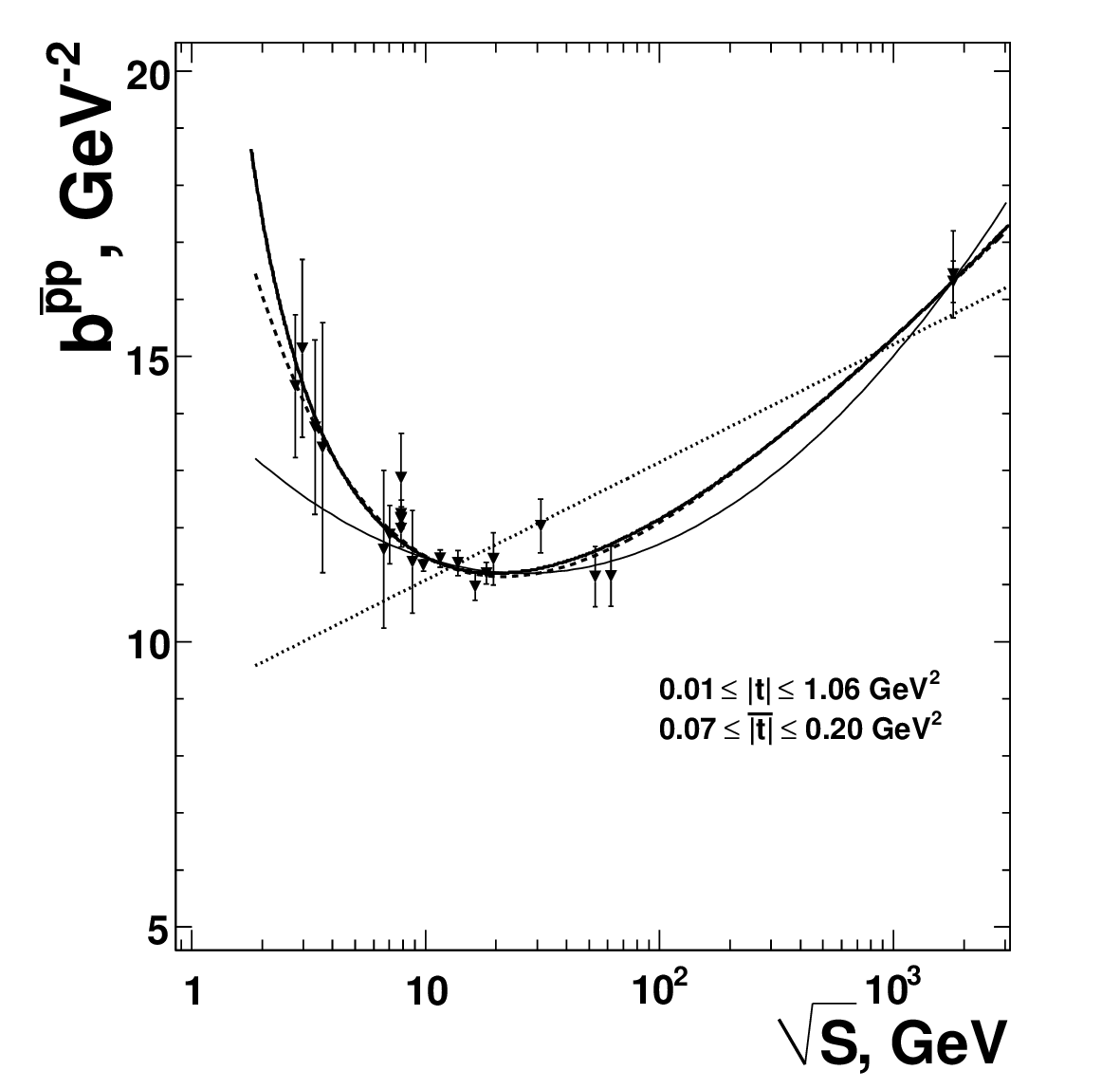}
} \caption{Energy dependence of the slope parameter $b$ at
$|t|=0.2$ GeV$^{2}$ for antiproton-proton scattering. The
correspondence of curves to the fit functions is the same as
previously}\label{fig:5}
\end{figure}

Experimental values of $b$ depend on energy collisions and
corresponding fits are shown at Fig.\ref{fig:4} for $pp$ elastic
scattering and at Fig.\ref{fig:5} for $\bar{p}p$ collisions. In
the last case fit qualities for (\ref{eq:Fit-1}), (\ref{eq:Fit-4})
functions are better for fitting at $\sqrt{s} \geq 5$ GeV only
than that for fitting of all available energy domain. The fit
parameter values are shown in Table \ref{tab:4}. Fit qualities are
better significantly than that for corresponding fits of $B$
parameter with the exception of (\ref{eq:Fit-1}) for $\bar{p}p$
data. Functions (\ref{eq:Fit-1}), (\ref{eq:Fit-4}) approximate
$b\left(\sqrt{s}\right)$ for $pp$ data statistically acceptable
for $\sqrt{s} \geq 5$ GeV only. Functions (\ref{eq:Fit-2}) and
(\ref{eq:Fit-3}) show acceptable close fit qualities and
difference at high energies only. The shrinkage parameter
$a_{1}^{pp}$ for best approximation function (\ref{eq:Fit-3}) is
in a good agreement with a early results
\cite{Burq-NPB-217-285-1983}. Function (\ref{eq:Fit-2}) shows a
best fit quality for $\bar{p}p$ data. Thus the "expanded"
parameterizations (\ref{eq:Fit-2}) and (\ref{eq:Fit-3}) suppose
statistically acceptable representation of all available
experimental data for $b$ parameter both in $pp$ and $\bar{p}p$
elastic reactions.

\begin{table*}
\caption{Fitting parameters for $b$ energy dependence}
\label{tab:4}
\begin{center}
\begin{tabular}{lccccc}
\hline \multicolumn{1}{l}{Function} &
\multicolumn{5}{c}{Parameter} \\
\cline{2-6} \rule{0pt}{10pt}
 & $B_{0}$, GeV$^{-2}$ & $a_{1}$, GeV$^{-2}$ & $a_{2}$, GeV$^{-2}$ & $a_{3}$ & $\chi^{2}/\mbox{n.d.f.}$ \\
\hline
\multicolumn{6}{c}{proton-proton scattering} \\
\hline
(\ref{eq:Fit-1}) & $6.7 \pm 0.2$ & $0.32 \pm 0.02$ & --              & --             & $25.6/25$ \\
(\ref{eq:Fit-2}) & $8.4 \pm 0.4$ & $0.20 \pm 0.03$ & $-10.4 \pm 1.4$ & $-2.0 \pm 4.5$ & $41.5/33$ \\
(\ref{eq:Fit-3}) & $7.6 \pm 0.8$ & $0.25 \pm 0.07$ & $-20 \pm 15$    & $-1.2 \pm 0.4$ & $41.4/33$ \\
(\ref{eq:Fit-4}) & $5.8 \pm 0.9$ & $0.5 \pm 0.2$   & $-0.05 \pm 0.04$& --             & $24.3/24$ \\
\hline
\multicolumn{6}{c}{proton-antiproton scattering} \\
\hline
(\ref{eq:Fit-1}) & $9.2 \pm 0.2$  & $0.22 \pm 0.02$ & --              & --               & $40.5/13$ \\
(\ref{eq:Fit-2}) & $-234 \pm 143$ & $0.71 \pm 0.07$ & $252 \pm 152$   & $-0.04 \pm 0.02$ & $8.8/15$ \\
(\ref{eq:Fit-3}) & $2.4 \pm 1.6$  & $0.46 \pm 0.11$ & $18.5 \pm 2.3$  & $-0.29 \pm 0.11$ & $9.0/15$ \\
(\ref{eq:Fit-4}) & $14 \pm 1$     & $-0.47 \pm 0.12$& $0.07 \pm 0.01$ & --               & $8.1/12$ \\
\hline
\end{tabular}
\end{center}
\vspace*{-0.4cm}
\end{table*}
\begin{table*}
\caption{Predictions for slope parameters based on the functions
(\ref{eq:Fit-1}) - (\ref{eq:Fit-4}) for intermediate $|t|$ domain
} \label{tab:5}
\begin{center}
\begin{tabular}{lccccccccccc}
\hline \multicolumn{1}{l}{Fitting} &
\multicolumn{11}{c}{Facility energies, $\sqrt{s}$} \\
\cline{2-12} \rule{0pt}{11pt} function & \multicolumn{4}{c}{FAIR,
GeV} & \multicolumn{2}{c}{NICA, GeV} &
\multicolumn{2}{c}{RHIC, TeV} & \multicolumn{3}{c}{LHC, TeV} \\
\cline{2-12} \rule{0pt}{10pt}
 & 3& 5 & 6.5 & 14.7 & 20 & 25 & 0.2 & 0.5 & 14 & 28 & 42$^*$ \\
\hline & \multicolumn{11}{c} {$B$-parameter} \\ \hline
(2a) & 11.80 & 12.06 & 12.19 & 12.60 & 10.15 & 10.32 & 11.85 & 12.52 & 14.98 & 15.45 & 15.79 \\
(2b) & 14.08 & 12.56 & 12.12 & 11.52 & 10.31 & 10.43 & 11.26 & 11.57 & 12.64 & 12.86 & 12.98 \\
(2c) & 13.29 & 12.52 & 12.19 & 11.50 & 10.23 & 10.36 & 11.56 & 12.09 & 14.01 & 14.41 & 14.64 \\
(2d) & 13.26 & 12.51 & 12.19 & 11.50 & 10.45 & 10.61 & 10.38 & 9.30 & 0.33 & -2.53 & -4.36 \\
\hline & \multicolumn{11}{c} {$b$-parameter} \\ \hline
(2a) & 10.00 & 10.46 & 10.70 & 11.43 & 10.52 & 10.81 & 13.44 & 14.60 & 18.81 & 19.68 & 20.20 \\
(2b) & 14.49 & 12.60 & 12.03 & 11.24 & 10.50 & 10.72 & 12.55 & 13.32 & 16.05 & 16.61 & 16.94 \\
(2c) & 14.22 & 12.65 & 12.10 & 11.22 & 10.49 & 10.72 & 12.79 & 13.69 & 16.99 & 17.67 & 18.08 \\
(2d) & 12.56 & 11.99 & 11.76 & 11.30 & 10.50 & 10.72 & 11.87 & 11.87 & 9.27  & 8.22  & 7.52 \\
\hline \multicolumn{10}{l}{$^*$\rule{0pt}{11pt}\footnotesize The
ultimate energy upgrade of LHC project \cite{Skrinsky-ICHEP2006}.}
\end{tabular}
\end{center}
\vspace*{-0.4cm}
\end{table*}
We have obtained predictions for nuclear slope parameters $B$ and
$b$ for some facilities based on the fit results shown above. The
predicted $B$ values at intermediate $|t|$ are calculated on the
base of fitting parameters obtained for linear exponential
parametrization of $d\sigma/dt$. Slope values are shown in the
Table \ref{tab:5} for different energies of FAIR, NICA, RHIC, and
LHC. According to the fit range function (\ref{eq:Fit-1}) can
predicts the $B$ value for $\bar{p}p$ scattering at all energies
under study not in $\sqrt{s} \geq 5$ GeV domain only. As expected
the functions (\ref{eq:Fit-3}) and (\ref{eq:Fit-4}) predicted very
close slope parameter values for FAIR. All fitting functions,
especially (\ref{eq:Fit-2}) and (\ref{eq:Fit-3}), predict the
close values for nuclear slope in NICA energy domain. Functions
(\ref{eq:Fit-1}) -- (\ref{eq:Fit-3}) predict larger values for $B$
in high-energy $pp$ collisions than (\ref{eq:Fit-4})
approximation. Perhaps, the future more precise RHIC results will
be useful for discrimination of fitting functions under study for
intermediate $|t|$ values. The function (\ref{eq:Fit-4}) with
obtained parameters predicts negative $B$ values at future LHC
energies. It should be emphasized that various phenomenological
models predict a very sharp decreasing of nuclear slope in the
range $|t| \sim 0.3 - 0.5$ GeV$^{2}$ at LHC energy $\sqrt{s}=14$
TeV [11]. Just the negative $B$ value predicted for LHC at
$\sqrt{s}=14$ TeV by (\ref{eq:Fit-4}) is most close to the some
model expectations
\cite{Petrov-EPJ-C28-525-2003,Bourrely-EPJ-C28-97-2003}. Taking
into account predictions in Table \ref{tab:2} based on the fitting
functions (\ref{eq:Fit-1}) -- (\ref{eq:Fit-4}) for low $|t|$ one
can suggest that the model with hadronic amplitude corresponding
to the exchange of three pomerons
\cite{Kundrat-EDS-273-2007,Petrov-EPJ-C28-525-2003} describes the
nuclear slope some closer to the experimentally inspired values at
LHC energy both at low and intermediate $|t|$ than other models.
One can see the functions (\ref{eq:Fit-2}) and (\ref{eq:Fit-3})
predict very close values of $b$ up to LHC energies. Function
(\ref{eq:Fit-4}) shows a much slower decreasing of $b$ at LHC
energy domain than that for $B$ parameter. The values of $b$
parameter differ significantly from each other for various fitting
functions in ultra-high energy domain for $pp$ collisions and at
low energies for $\bar{p}p$ elastic scattering. It seems $b$
parameter is more perspective for distinguishing of Regge-inspired
function (\ref{eq:Fit-1}) from "expanded" parameterizations
(\ref{eq:Fit-2}), (\ref{eq:Fit-3}) at $\sqrt{s} \geq 0.5$ TeV than
$B$ because of larger differences between predictions of $b$ for
corresponding functions.
\subsection{$\Delta B$ and $NN$ data analysis}
\begin{figure*}
\begin{center}
\begin{tabular}{cc}
\mbox{
\resizebox{0.5\textwidth}{!}{%
  \includegraphics[width=8.0cm,height=8.0cm]{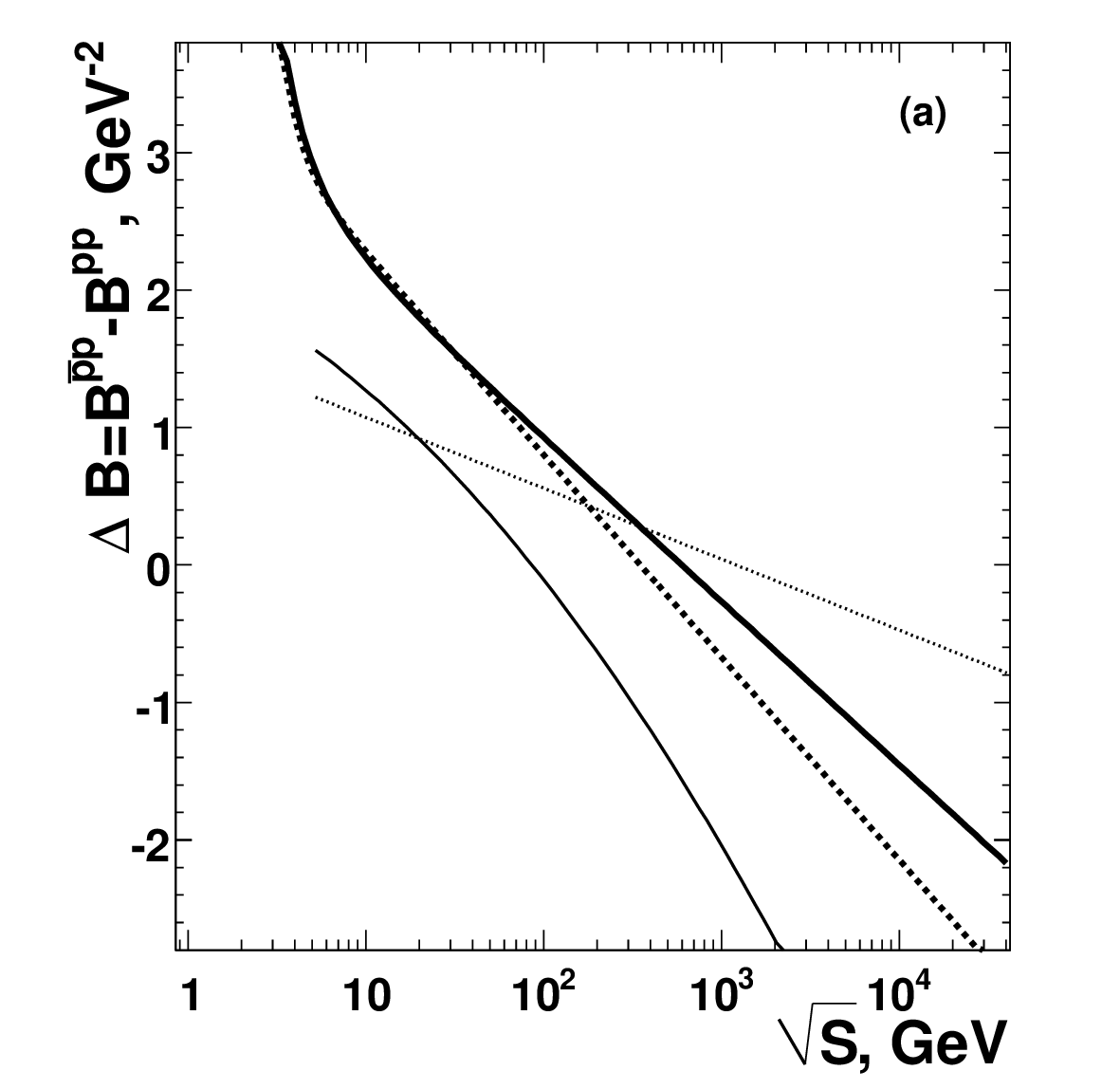}
}} & \mbox{
\resizebox{0.5\textwidth}{!}{%
  \includegraphics[width=8.0cm,height=8.0cm]{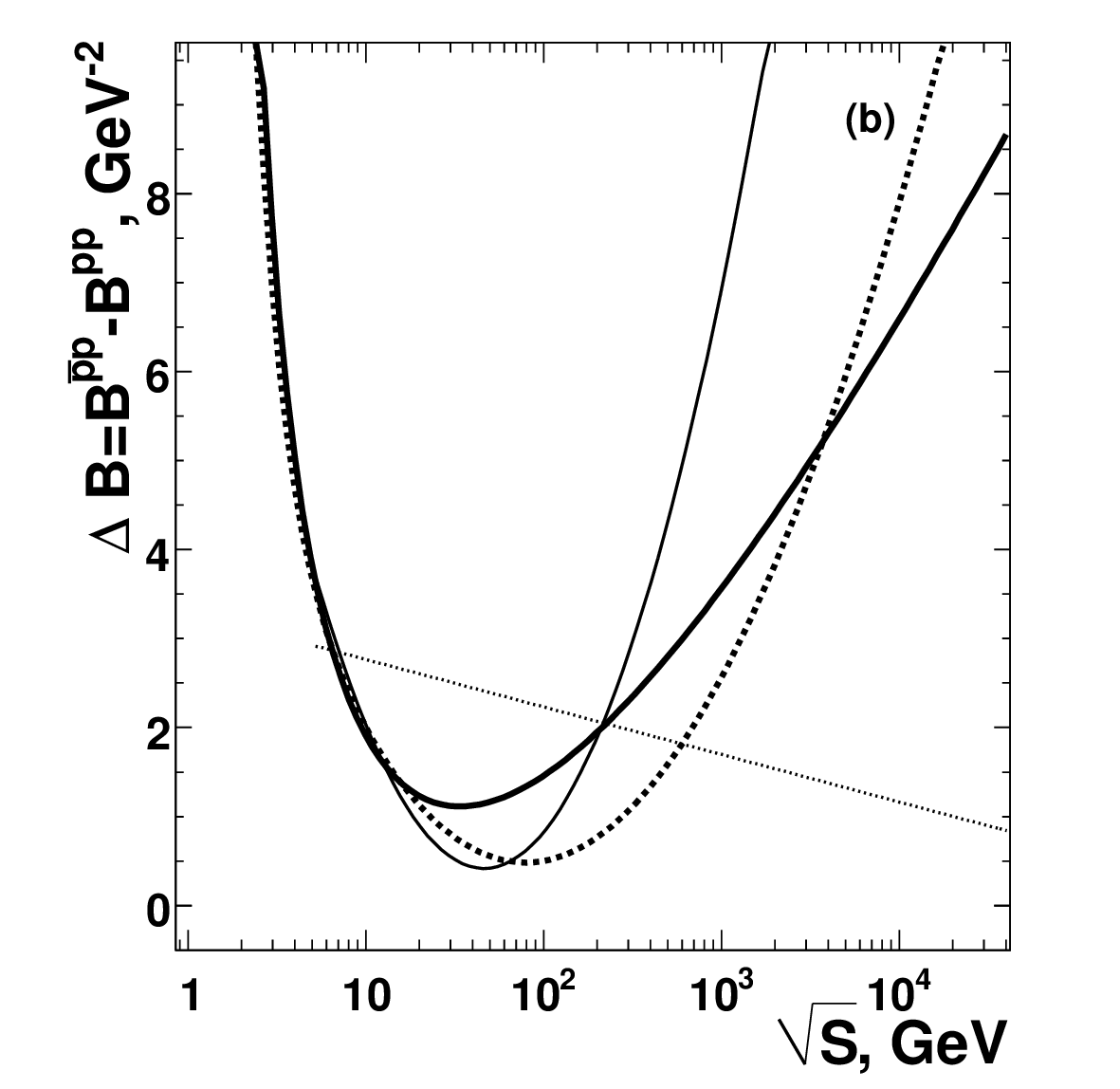}
}}
\end{tabular}
\end{center}
\vspace*{-0.4cm} \caption{The energy dependence of the difference
of elastic slopes for proton-antiproton and proton-proton
scattering in low $|t|$ domain (a) and in intermediate $|t|$ range
for linear exponential fit of differential cross-section (b). The
correspondence of curves to the fit functions is the same as
above}\label{fig:6}
\end{figure*}

Phenomenological models predicts the zero difference of slopes
$\left(\Delta B\right)$ for proton-antiproton and proton-proton
elastic scattering at asymptotic energies. Here the difference
$\Delta B$ is calculated for each function (\ref{eq:Fit-1}) --
(\ref{eq:Fit-4}) under study with parameters corresponded
$\bar{p}p$ and $pp$ fits: $\Delta
B_{i}\left(s\right)=B^{\bar{p}p}_{i}\left(s\right)-B^{pp}_{i}
\left(s\right),~i=\mbox{2a,...2d}$\footnote{Obviously, one can
suggest various combinations of fitting functions for $\Delta B$
calculations, for example, the difference between fitting
functions with best fit qualities etc.}. It should be stressed
that the equal energy domain are used in $\bar{p}p$ and $pp$ fits
for $\Delta B$ calculations, i.e. the parameter values obtained by
(\ref{eq:Fit-4}) fitting function for $\bar{p}p$ data from linear
exponential fit of $d\sigma/dt$ for $\sqrt{s} \geq 5$ GeV are used
for corresponding $\Delta B$ definition. The energy dependence of
$\Delta B$ is shown at Fig.\ref{fig:6}a and Fig.\ref{fig:6}b for
low and intermediate $|t|$ respectively. One can see that the
difference of slopes decreasing with increasing of energy for low
$|t|$ domain (Fig.\ref{fig:6}a). At present the proton-proton
experimental data at highest available energy 200 GeV don't
contradict with fast (square of logarithm of energy) increasing of
slope at high energies in general case. Such behavior could be
agreed with the asymptotic growth of total cross section. But on
the other hand the quadratic logarithmic function (\ref{eq:Fit-4})
leads to very significant difference $\Delta B$ for $\bar{p}p$ and
$pp$ scattering in high energy domain for both low
(Fig.\ref{fig:6}a) and intermediate (Fig.\ref{fig:6}b) values of
$|t|$. The only Regge-inspired function (\ref{eq:Fit-1}) predicts
the decreasing of $\Delta B$ with energy growth at intermediate
$|t|$ (Fig.\ref{fig:6}b). The parameterizations (\ref{eq:Fit-2})
-- (\ref{eq:Fit-4}) predict the decreasing of difference of slopes
at low and intermediate energies and fast increasing of $\Delta B$
at higher energies for intermediate $|t|$ domain
(Fig.\ref{fig:6}b). As expected the most slow changing of $\Delta
B$ is predicted by Regge-inspired function (\ref{eq:Fit-1}) at
asymptotic energies. All fitting functions with experimentally
inspired parameters don't predict the constant zero values of
$\Delta B$ at high energies. But it should be emphasized that only
separate fits were made for experimental data for $pp$ and
$\bar{p}p$ elastic reactions above. These results indicate on the
importance of investigations at ultra-high energies both $pp$ and
$\bar{p}p$ elastic scattering for many fundamental questions and
predictions related to the general asymptotic properties of
hadronic physics.

\begin{figure}
\resizebox{0.5\textwidth}{!}{%
  \includegraphics[width=8.0cm,height=8.0cm]{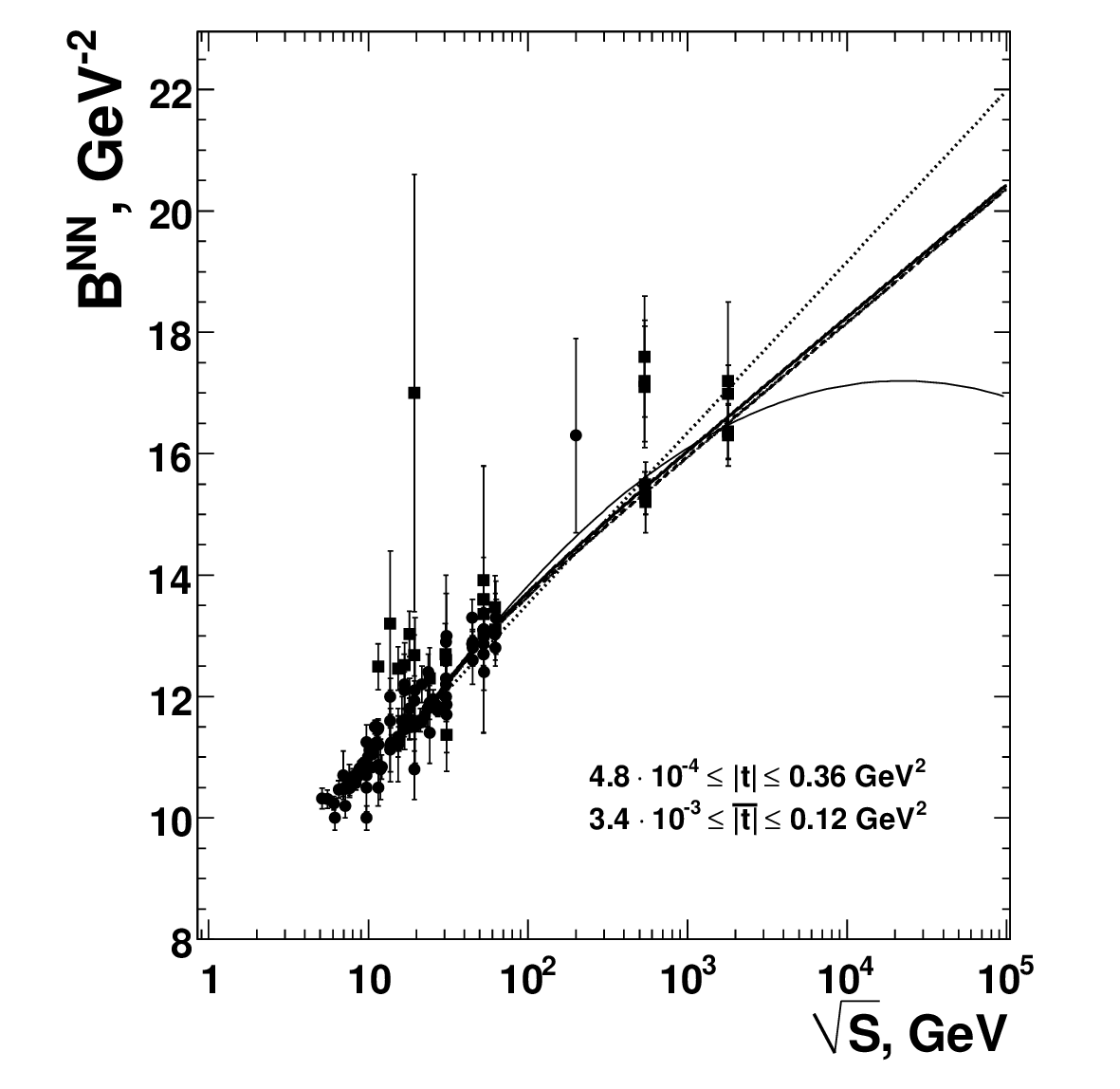}
} \caption{Energy dependence of the elastic slope parameter for
nucleon-nucleon scattering for low $|t|$ domain. Experimental
fitted points are indicated as {\large$\bullet$} for $pp$ and as
$\blacksquare$ for $\bar{p}p$. Fits are shown for
$\sqrt{s_{\mbox{\small{min}}}}=20$ GeV. The correspondence of
curves to the fit functions is the same as previously}
\label{fig:7}
\end{figure}

\begin{figure}
\resizebox{0.5\textwidth}{!}{%
  \includegraphics[width=8.0cm,height=8.0cm]{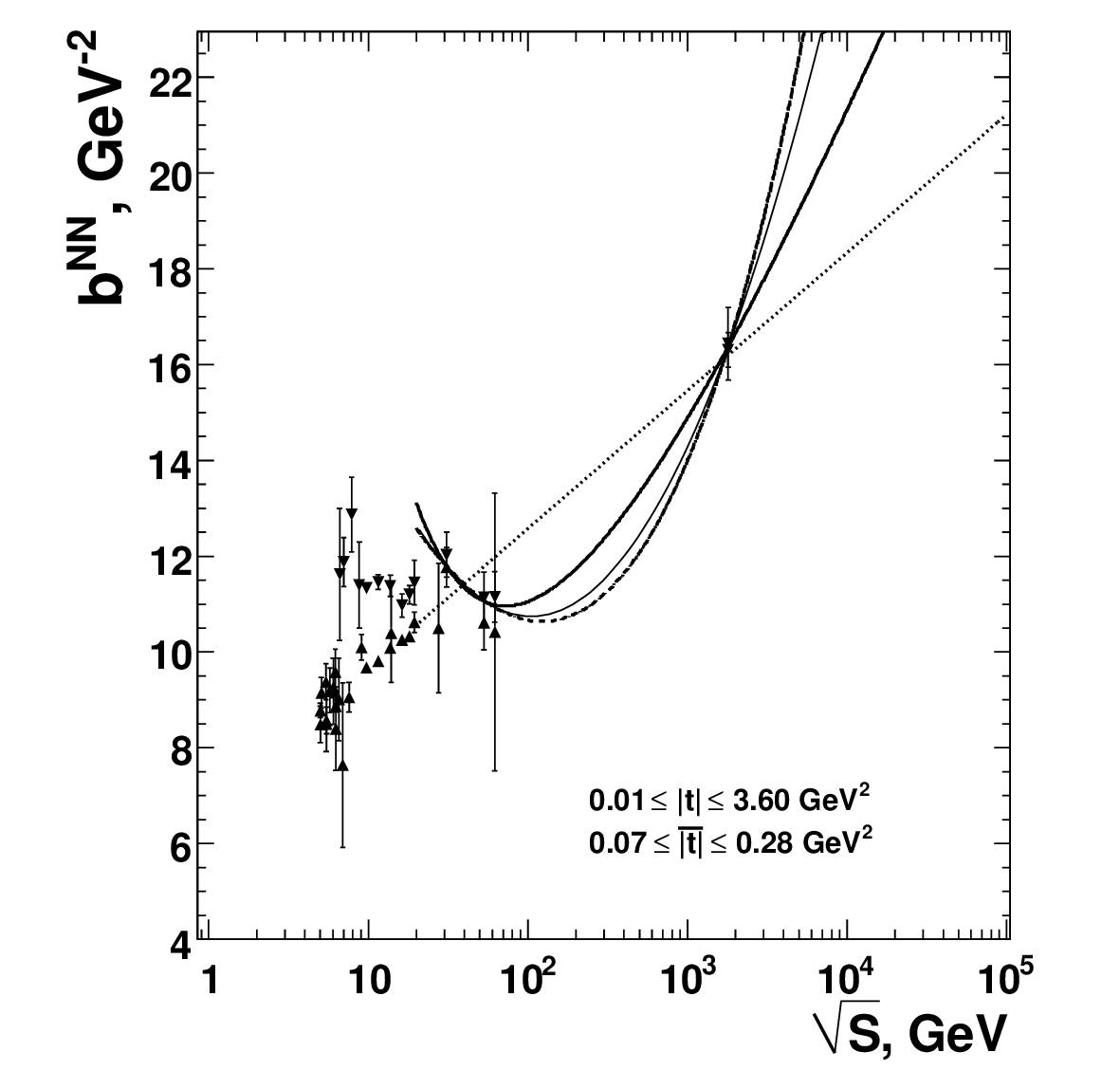}
} \caption{Energy dependence of the elastic slope parameter $b$ at
$|t|=0.2$ GeV$^{2}$ for nucleon-nucleon scattering for
intermediate $|t|$ domain. Experimental fitted points are
indicated as $\blacktriangle$ for $pp$ and as $\blacktriangledown$
for $\bar{p}p$. Fits are shown for
$\sqrt{s_{\mbox{\small{min}}}}=20$ GeV. The correspondence of
curves to the fit functions is the same as previously}
\label{fig:8}
\end{figure}

Also we have analyzed general data samples for $pp$ and $\bar{p}p$
elastic scattering. Slope parameters ($B$ and $b$) shows a
different energy dependence at $\sqrt{s} < 5$ GeV in proton-proton
and antiproton-proton elastic reactions in any $|t|$ domains under
study. Thus slopes for nucleon-nucleon data are investigated only
for $\sqrt{s} \geq 5$ GeV below. We have included in fitted
samples only $pp$ and $\bar{p}p$ points which were included in
corresponding final data samples at separate study $pp$ and
$\bar{p}p$ elastic reactions above. We did not exclude any points
from $NN$ data sample, we change only the low energy boundary for
fitted domain. Fig.\ref{fig:7} shows the experimental data for
slope in nucleon-nucleon elastic scattering against collision
energy at low $|t|$. As seen from Fig.\ref{fig:7} there is no
experimental data for $\bar{p}p$ between $\sqrt{s}=5$ GeV and
$\sqrt{s}=10$ GeV.
This energy domain will available for further FAIR facility. We
have fitted the general nucleon-nucleon data sample at range of
low energy boundary $s_{\mbox{\small{min}}}=25-400$ GeV$^{2}$. The
fitting parameter values are indicated in Table \ref{tab:6} on the
first line for $s_{\mbox{\small{min}}}=25$ GeV$^{2}$ and on the
second one for $s_{\mbox{\small{min}}}=400$ GeV$^{2}$. The fit
quality improves for all parameterizations under consideration at
increasing of $s_{\mbox{\small{min}}}$. One need to emphasize the
fit quality is some poorer $\left(\chi^{2}/\mbox{n.d.f.} \! \simeq
\! 2.3-2.4\right)$ at $\sqrt{s} \! \geq \! 10$ GeV than that for
$\sqrt{s} \! \geq \! 5$ GeV for functions (\ref{eq:Fit-1}) and
(\ref{eq:Fit-4}). In the first case the $a_{1}$ parameter value is
close to the Regge model prediction within errors for function
(\ref{eq:Fit-4}) only. The value $a_{1}=0.24 \pm 0.01$ GeV$^{-2}$
obtained for fitting functions (\ref{eq:Fit-2}) and
(\ref{eq:Fit-3}) at $s_{\mbox{\small{min}}}=400$ GeV$^{2}$ agrees
very well with estimate for asymptotic shrinkage parameter
$\alpha'_{\cal{P}}$ . All functions (\ref{eq:Fit-1}) --
(\ref{eq:Fit-4}) are close to each other at energies up to
$\sqrt{s} \! \sim \! 10$ TeV at least and shows quasi-linear
behavior for parameter values obtained by fits with
$s_{\mbox{\small{min}}}=25$ GeV$^{2}$. Fitting functions
(\ref{eq:Fit-1}) -- (\ref{eq:Fit-4}) are shown at Fig.\ref{fig:7}
for $s_{\mbox{\small{min}}}=400$ GeV$^{2}$. The function
(\ref{eq:Fit-4}) decreases at ultra-high energies $\sqrt{s} \! >
\! 20$ TeV and clear separation of various fitting functions is
accessible in the LHC energy domain. We have analyzed the
nucleon-nucleon data for slope parameters $B$ and $b$ at
intermediate $|t|$ values for linear and quadratic exponential
parametrization of $d\sigma / dt$ respectively. Fit results are
shown in Table \ref{tab:7}. Experimental $pp$ and $\bar{p}p$ data
for $B$ differ significantly up to $\sqrt{s} \simeq 10$ GeV at
least that results in unacceptable fit qualities for all functions
under study ($\chi^{2}/\mbox{n.d.f.} \simeq 29.6$ for best fit by
quadratic logarithmic function). Additional analysis demonstrate
the improving of fit quality for (\ref{eq:Fit-2}) --
(\ref{eq:Fit-4}) with increasing of low energy boundary from
$s_{\mbox{\small{min}}}=25$ GeV$^{2}$ up to
$s_{\mbox{\small{min}}}=400$ GeV$^{2}$. The values of fit
parameters for the last case are shown in Table \ref{tab:7}. The
Regge-inspired function (\ref{eq:Fit-1}) contradicts with
experimental data. We would like to emphasize that the best fit
quality for (\ref{eq:Fit-1}) is obtained at
$s_{\mbox{\small{min}}}=100$ GeV$^{2}$
$\left(\chi^{2}/\mbox{n.d.f.} \! \simeq \! 7.7\right)$ but it is
statistically unacceptable too. Functions (\ref{eq:Fit-2}) --
(\ref{eq:Fit-4}) represent experimental $B\left(\sqrt{s}\right)$
reasonably and have a very close fit qualities. One can see from
Table \ref{tab:7} the statistically acceptable fits have been
obtained for $b$ parameter at $s_{\mbox{\small{min}}}=400$
GeV$^{2}$ only. Experimental data and fit functions are presented
at the Fig.\ref{fig:8}. Functions (\ref{eq:Fit-2}) --
(\ref{eq:Fit-4}) show close fit qualities. Best fit is
(\ref{eq:Fit-4}) but "expanded" parameterizations agree with data
too. One needs to emphasize the significant errors and absence of
experimental points at $\sqrt{s} \simeq 0.1-2$ TeV that trouble
the more clear conclusion. The RHIC as well as LHC data for
nucleon-nucleon differential cross-section at intermediate $|t|$
will be helpful for distinguishing of various fit functions.

One can conclude the slope parameters for $pp$ and $\bar{p}p$
elastic scattering show universal behavior at $\sqrt{s} \geq 20$
GeV and "expanded" functions represent the energy dependencies for
both low and intermediate $|t|$ ranges for this energy domain.
Thus quantitative analysis of slopes at different $|t|$ allows us
to get the following estimation of low energy boundary: $\sqrt{s}
\simeq 20$ GeV for universality of elastic nucleon-nucleon
scattering. This estimates agrees with results for differential
cross-sections of $pp$ and $\bar{p}p$ elastic reactions based on
the crossing-symmetry and derivative relations
\cite{Okorokov-arXiv-0711.2231}.
\begin{table*}
\caption{Fitting parameters for slope energy dependence in low
$|t|$ domain for $NN$ elastic scattering} \label{tab:6}
\begin{center}
\begin{tabular}{lccccc}
\hline \multicolumn{1}{l}{Function} &
\multicolumn{5}{c}{Parameter} \\
\cline{2-6} \rule{0pt}{10pt}
 & $B_{0}$, GeV$^{-2}$ & $a_{1}$, GeV$^{-2}$ & $a_{2}$, GeV$^{-2}$ & $a_{3}$ & $\chi^{2}/\mbox{n.d.f.}$ \\
\hline
(\ref{eq:Fit-1}) & $8.35 \pm 0.06$ & $0.278 \pm 0.005$ & --                 & --               & $272/124$ \\
     & $7.89 \pm 0.11$ & $0.306 \pm 0.008$ & --                 & --               & $82.8/60$ \\
(\ref{eq:Fit-2}) & $8.21 \pm 0.07$ & $0.285 \pm 0.005$ & $1.6 \pm 0.4$      & $-2.0 \pm 2.4$   & $271/122$ \\
     & $9.54 \pm 0.25$ & $0.24 \pm 0.01$   & $-514 \pm 244$     & $-3.5 \pm 0.3$   & $59.1/58$ \\
(\ref{eq:Fit-3}) & $8.05 \pm 0.55$ & $0.29 \pm 0.02$   & $0.64 \pm 0.55$    & $-0.3 \pm 0.4$   & $270/122$ \\
     & $9.3 \pm 0.2$   & $0.24 \pm 0.01$   & $-43 \pm 18$       & $-0.67 \pm 0.08$ & $59.1/58$ \\
(\ref{eq:Fit-4}) & $8.48 \pm 0.14$ & $0.26 \pm 0.02$   & $\left(3 \pm 3\right) \cdot 10^{-3}$  & -- & $271/123$ \\
     & $5.6 \pm 0.5$   & $0.58 \pm 0.06$   & $-0.029 \pm 0.006$ & -- & $61.1/59$ \\
\hline
\end{tabular}
\end{center}
\end{table*}

\begin{table*}
\caption{Fitting parameters for energy dependence of slope
parameters at intermediate $|t|$ for $NN$ elastic scattering}
\label{tab:7}
\begin{center}
\begin{tabular}{lccccc}
\hline \multicolumn{1}{l}{Function} &
\multicolumn{5}{c}{Parameter} \\
\cline{2-6} \rule{0pt}{10pt}
 & $B_{0}$, GeV$^{-2}$ & $a_{1}$, GeV$^{-2}$ & $a_{2}$, GeV$^{-2}$ & $a_{3}$ & $\chi^{2}/\mbox{n.d.f.}$ \\
\hline & \multicolumn{5}{c} {$B$-parameter} \\ \hline
(\ref{eq:Fit-1}) & $5.42 \pm 0.12$& $0.353 \pm 0.007$& --               & --                 & $271/34$ \\
(\ref{eq:Fit-2}) & $-253 \pm 44$  & $1.00 \pm 0.06$  & $277 \pm 46$     & $-0.06 \pm 0.01$   & $110/32$ \\
(\ref{eq:Fit-3}) & $-162 \pm 28$  & $2.5 \pm 0.2$    & $177 \pm 28$     & $-0.036 \pm 0.003$ & $103/32$ \\
(\ref{eq:Fit-4}) & $14.1 \pm 0.7$ & $-0.53 \pm 0.07$ & $0.081 \pm 0.006$& -- & $102/33$ \\
\hline & \multicolumn{5}{c} {$b$-parameter} \\ \hline
(\ref{eq:Fit-1}) & $6.97 \pm 0.13$& $0.318 \pm 0.012$  & --                 & --                 & $221/40$ \\
     & $6.8 \pm 0.5$  & $0.31 \pm 0.03$    & --                 & --                 & $15.3/7$ \\
(\ref{eq:Fit-2}) & $7.5 \pm 0.2$  & $0.282 \pm 0.017$  & $-595 \pm 662$     & $-6 \pm 3$ & $206/38$ \\
     & $-23 \pm 8$    & $0.98 \pm 0.19$    & $144 \pm 40$       & $-1 \pm 2$ & $2.43/5$ \\
(\ref{eq:Fit-3}) & $7.5 \pm 0.2$  & $0.281 \pm 0.016$  & $-152 \pm 287$     & $-1.7 \pm 0.6$   & $205/38$ \\
     & $9.0 \pm 28.5$ & $-1.7 \pm 1.4$     & $13 \pm 31$        & $0.10 \pm 0.08$  & $2.40/5$ \\
(\ref{eq:Fit-4}) & $6.5 \pm 0.3$  & $0.40 \pm 0.04$    & $-0.010 \pm 0.005$  & -- & $217/39$ \\
     & $26 \pm 5$     & $-1.6 \pm 0.5$     & $0.17 \pm 0.05$ & -- & $2.40/6$ \\
\hline
\end{tabular}
\end{center}
\end{table*}

\section{Conclusions}
\label{sec:3}The main results of this paper are the following.
Slope energy dependencies are analyzed quantitatively for elastic
nucleon-nucleon scattering in various $|t|$ domains. Most of all
available experimental data for slope parameter in elastic nucleon
collisions are approximated by different analytic functions.

The suggested new parameterizations allow us to describe
experimental nuclear slope at all available energies in low $|t|$
domain for $pp$ quite reasonably. The new approximations agree
with experimental $\bar{p}p$ data at qualitative level but these
fits are still statistically unacceptable because of very sharp
behavior of $B$ near the low energy limit. The best fit quality is
obtained for "expanded" logarithmic function both for $pp$ and
$\bar{p}p$ data. The obtained values of asymptotic shrinkage
parameter $\alpha'_{\cal{P}}$ for $pp$ elastic scattering are
larger than $\alpha'_{\cal{P}}$ values for elastic $\bar{p}p$
reactions for the same fitting functions. Various approximations
differ from each other both in the low energy and very high energy
domains. Predictions for slope parameter are obtained for elastic
proton-proton and proton-antiproton scattering in energy domains
of some facilities at low momentum transfer. Our predictions based
on the all available experimental data don't contradict the
phenomenological model estimations qualitatively. The situation is
more unclear at intermediate $|t|$ values than for low $|t|$
domain. Only the qualitative agreement is observed between
approximations and experimental points both for $pp$ and
$\bar{p}p$ collisions for linear exponential parametrization of
$d\sigma /dt$ because of poorer quality of data. The "expanded"
logarithmic function describes of $pp$ data for $B$ parameter for
any differential cross-section parametrization reasonably. Best
fit quality is obtained for quadratic function of logarithm for
$\bar{p}p$ data. One needs to emphasize that this function allows
us to describe $\bar{p}p$ data at all available energies and shows
a statistically acceptable fit quality for data sample obtained
from quadratic exponential parametrization of $d\sigma / dt$.
Slope parameter $b$ calculated at $|t|=0.2$ GeV$^{2}$ shows more
smooth energy dependence. We have obtained an acceptable fit
qualities for "expanded" functions both for $pp$ and $\bar{p}p$
data at all initial energies. The obtained values of asymptotic
shrinkage parameter $\alpha'_{\cal{P}}$ for $pp$ elastic
scattering are close to the early results both in low and
intermediate $|t|$ domain. As well as for low $|t|$ domain
predictions for slope parameters $B$ and $b$ are obtained for
elastic proton-proton and proton-antiproton scattering in energy
domains of some facilities. It seems the phenomenological model
with hadronic amplitude corresponding to the exchange of three
pomerons describes the nuclear slope some closer to the
experimental fit inspired values at LHC energy both at low and
intermediate $|t|$ than other models.

The energy dependence of difference of slopes $\left(\Delta
B\right)$ for proton-antiproton and proton-proton elastic
scattering was obtained for fitting functions under study. The
$\Delta B$ parameter shows the opposite behaviors at high energies
for low and intermediate $|t|$ domains (decreasing / increasing,
respectively) for all fitting functions with the exception of
Regge-inspired one. The last function predicts the slow decreasing
of $\Delta B$ with energy growth. It should be emphasized that all
underlying empirical fitting functions with experimentally
inspired parameter values don't predict the zero difference of
slopes for proton-antiproton and proton-proton elastic scattering
both at low and intermediate $|t|$ for high energy domain.
We have analyzed general nucleon-nucleon data samples for slopes
at $\sqrt{s} \geq 5$ GeV. The "expanded" functions show the best
and statistically acceptable fit qualities at $\sqrt{s} \geq 20$
GeV for low $|t|$ domain. Slop analysis allows us to find the
following value $0.24 \pm 0.01$ for $\alpha'_{\cal{P}}$ parameter.
The estimation of asymptotic shrinkage parameter
$\alpha'_{\cal{P}}$ obtained with "expanded" functions for $NN$
data agree very well with Pomeron theory expectation. The
quadratic logarithmic function represents experimental $NN$ data
for $B$ and $b$ slope parameters with best quality. But the
functions (\ref{eq:Fit-2}) and (\ref{eq:Fit-3}) show a close
qualities and agree with data reasonably. Therefore suggested
"expanded" functions can be used as a reliable fits for wide range
of momentum transfer at all energies. The universal behavior was
found for available experimental $pp$ and $\bar{p}p$ slopes at
$\sqrt{s} \geq 20$ GeV both for low and intermediate $|t|$ that is
in agreement with the hypothesis of a universal shrinkage of the
hadronic diffraction cone at high energies.

\end{document}